\documentclass[manuscript,revised]{geophysics}
\usepackage{comment}

\usepackage{bm}
\usepackage{textcomp}
\usepackage[T1]{fontenc}
\usepackage{amsmath}
\newcommand{\overbar}[1]{\mkern 1.5mu\overline{\mkern-1.5mu#1\mkern-1.5mu}\mkern 1.5mu}
\usepackage{amssymb}
\usepackage{mathtools}  
\usepackage{amsfonts}
\usepackage{xcolor}
\usepackage[colorlinks=true,citecolor=blue,linkcolor=blue,urlcolor=red]{hyperref}
\usepackage{float}
\usepackage[font=small,skip=0pt]{caption}
\usepackage[version=4]{mhchem}

\usepackage[left= 2.6cm, right=2.6cm, top =2.6cm, bottom=2.6cm]{geometry}
\begin{document}

\title{A Bayesian-based approach to improving acoustic Born waveform inversion of seismic data for viscoelastic media}

\renewcommand{\thefootnote}{\arabic{footnote}}

\address{
\footnotemark[1]Department of Applied Physics, University of Eastern Finland,
Kuopio, Finland

\footnotemark[2]School of Engineering Science, LUT University, Lappeenranta, Finland

\footnotemark[3]Nokia Bell Laboratories, Espoo, Finland}

\author{Kenneth Muhumuza\footnotemark[1], Lassi Roininen\footnotemark[2], Janne M. J. Huttunen\footnotemark[3], and Timo L{\"a}hivaara\footnotemark[1]}

\righthead{Acoustic waveform inversion of viscoelastic data}

\maketitle

\begin{abstract}

In seismic waveform inversion, the reconstruction of the subsurface properties is usually carried out using approximative wave propagation models to ensure computational efficiency. The viscoelastic nature of the subsurface is often unaccounted for, and two popular approximations---the acoustic and linearized Born inversion---are widely used. This leads to reconstruction errors since the approximations ignore realistic (physical) aspects of seismic wave propagation in the heterogeneous earth. In this study, we show that the Bayesian approximation error approach can be used to partially recover from errors, addressing elastic and viscous effects in acoustic Born inversion for viscoelastic media. The results of numerical examples indicate that neglecting the modeling errors induced by the approximations results in very poor recovery of the subsurface velocity fields.

\end{abstract}

\section{Introduction}
In geophysics, seismic waveform inversion is often used to obtain quantitative estimates of subsurface properties which predict the observed seismic data. The reconstruction of subsurface properties is a highly non-linear inverse problem. There are several techniques to solve the seismic inverse problem, for example, migration-based traveltime approaches \citep{zelt1992seismic,luo2016full}, Born approximation \citep{hudson1981use,kazei2018waveform,muhumuza2018seismic}, and full-waveform inversion approaches \citep{warner2013anisotropic,jakobsen2015full,virieux2017introduction}. The traveltime inversion is an approximation to the wave equation based on ray theory and uses traveltime information only. The Born approximation  corresponds to linearizing the nonlinear relationship between the data and the subsurface properties with the single-scattering approximation. The full-waveform inversion uses the full information content, including
both the amplitude and the traveltime of the recorded seismic data. Although ray-based and Born-based approximations are computationally cheaper than the desirable full-waveform methods, underlying assumptions may cause inaccuracies of their results.  

The core of a successful waveform inversion is an accurate forward solver, which honors both the dynamics and kinematics of seismic waves by accounting for the anisotropic and viscoelastic nature of the earth. However, accurate solvers are not often applied due to the high computational cost of solving a multidimensional nonlinear minimization problem. In practice, the viscoelastic nature of the subsurface is often disregarded, and several assumptions and approximate solutions of the wave equation are used to make computations more efficient. There are two widely used approximations which make seismic waveform inversion tractable --- The acoustic approximation, where only P-waves are modeled \citep[e.g.,][]{warner2013anisotropic,jakobsen2015full,masmoudi2018full}, and linearized Born (single scattering approximation) inversion  \citep{bleistein1985extension,symes2008approximate,weglein2009clarifying,jakobsen2015full,kazei2018waveform}. These approximations have consequences because some important aspects of the real-world physics are ignored. For example, while acoustic waveform inversion can account for the correct kinematics of the waves, it does not ideally account for P-S (and S-P) mode-converted reflections in a layered earth. This implies that the amplitudes for the elastic P-waves are incorrectly modeled due to the elastic effects \citep[e.g.,][]{chapman2014correcting,cance2015validity}. Also, methods based on linearization of the inverse problem using the Born approximation are not congruent with seismic wave propagation mechanics in a strongly scattering (heterogeneous) medium \citep[e.g.,][]{parisi2014validity,chen2015full,malovichko2017approximate,kaipio2019bayesian}. 

Several papers have been published that discuss the problems and consequences of implementing acoustic inversion using elastic data on the P-wave model reconstruction \citep{barnes2009domain,he2017analysis,mora2018elastic}. Some strategies have been suggested to address these problems in waveform inversion \citep{mulder2008exploring,hobro2014method,agudo2018acoustic}. Additionally,  \cite{coates1991generalized}, \cite{weglein2003inverse}, \cite{wu2014non}, and others have studied the limitations of Born approximation-based modeling and inversion methods in seismic exploration applications and found that nonlinear effects in wave propagation cannot be ignored. A great deal of work has been done to improve the convergence of the inverse Born series and overcome the limitations of Born approximation to extend the validity of Born-based waveform inversion \citep{jakobsen2012t,ouyang2014seismic,hou2015approximate,wang2017accurate,zuberi2017mitigating}. 

In this paper, we consider the seismic inverse problem in the Bayesian inversion framework \citep{tarantola2005inverse,kaipio2006statistical,bui2013computational,azevedo2017geostatistical}. This probabilistic framework provides, in principle, a methodology for incorporating parameter errors into the inversion, giving feasible (realistic) estimates with a measure of uncertainty. The posterior distribution can be mathematically interpreted as the solution of the inverse problem, but typically point estimates are needed for a practical solution.
Furthermore, we can assess the reliability of the reconstructed models by tabulating the Bayesian credible intervals of the estimated parameter. 

In order to partially dispense with the assumptions of the acoustic and Born approximations, we adopt the Bayesian approximation error (BAE) approach \citep{kaipio2006statistical,kaipio2007statistical}. The BAE approach uses two models: the accurate model and the approximative model. The first is used only to carry out offline precomputations; the second is used both in the precomputation stage and in the computation of estimates when data is measured. This particular approach thus takes into account a vast number of uncertainties in the forward model. In fact, the questions studied in this paper are: (1) what are the consequences on the waveform inversion results of using Born (and acoustic) approximations in an elastic world, at least under ideal assumptions? and (2) how well can one recover from the above consequences (and errors) using the BAE approach? We further investigate the impact on the reconstruction of the velocity field of using acoustic Born inversion in the presence of attenuation. 

The Bayesian approximation error approach was originally used to take into account the modeling errors induced by numerical model reduction \citep{kaipio2006statistical,kaipio2007statistical}. It is still used as such today, but it is also applied to handle various approximation and modeling errors in wide variety of inverse problems \citep{lehikoinen2007approximation,nissinen2007bayesian,nissinen2011reconstruction,kaipio2013approximate,koponen2014bayesian,lahivaara2015estimation,mozumder2016approximate,nicholson2018estimation}. Once we specify our approximate models, in the BAE, any errors induced by the use of simplified models, reducing the dimension of the parameter space, and/or model uncertainties are embedded into a single additive error term. 

Our goal is to study, for simple and easy to understand test examples, the kind of artefacts that may occur when applying the highly approximative acoustic Born waveform inversion to viscoelastic data, and whether BAE approach is able to recover from the errors. Using the BAE approach may make it possible to replace an accurate physical model that is computationally demanding with a less accurate but computationally feasible model. 

The rest of the paper is structured as follows: In Section \ref{Section2} we discuss wave propagation in isotropic elastic/acoustic media and introduce the Born approximation. The formulation and solution of the forward problem is presented in Section \ref{Section3}. In Section \ref{Section4} we give a brief overview of the Bayesian approach to inverse problems, introduce the Bayesian approximation error approach, and derive the estimators in the general form. In Section \ref{Section5} we consider two test cases and discuss the results. Finally, the conclusion is given in Section \ref{Section6}.

\section*{Governing equations for wave propagation} \label{Section2}

\subsection*{Wave propagation in elastic media}
Consider an isotropic material in which the material properties at a given point are identical in all directions. A seismic wave propagating in an elastic domain $\mathbf{\Omega }$ is governed by the following frequency-domain elastic wave equations  
\begin{align}
-\omega ^{2}\rho (\mathbf{x})\mathbf{u}(\mathbf{x},\omega )  & = \nabla \cdot \boldsymbol{\sigma }(\mathbf{u}) + \mathbf{f}(\mathbf{x},\omega ) \label{eqn1} \\
  \boldsymbol{\sigma }(\mathbf{x},\omega )  & = \mathbf{C}(\mathbf{x})\boldsymbol{\varepsilon}(\mathbf{x},\omega ),
\label{eqn2}
\end{align}
where $\omega$ is the angular frequency, $\mathbf{x}\in\mathbf{\Omega }$ is the position vector, $\rho (\mathbf{x})$ is the density, $\mathbf{u}(\mathbf{x},\omega)$ is the displacement vector, $\boldsymbol{\sigma }(\mathbf{x},\omega )$ is the stress tensor, $\mathbf{f}(\mathbf{x},\omega )$ is the source term,  $\mathbf{C}(\mathbf{x})$ is the elastic stiffness tensor, and $\boldsymbol{\varepsilon} (\mathbf{x},\omega ) = (\nabla\mathbf{u}+(\nabla\mathbf{u})^{T})/2$ is the strain tensor. For the isotropic case, $\mathbf{C}(\mathbf{x})$ depends upon only two Lam\'{e} parameters $\lambda (\mathbf{x})$ and  $\mu (\mathbf{x})$, such that stress tensor (\ref{eqn2}) can be written in the form 
\begin{equation}
 \boldsymbol{\sigma }(\mathbf{x},\omega ) = \lambda(\nabla\cdot\mathbf{u})\textup{I} + 2\mu \boldsymbol{\varepsilon}(\mathbf{x},\omega ) , 
\label{eqn4}
\end{equation} 
where $\textup{I}$ is the identity matrix. Isotropic P-wave and S-wave velocities $V_{p}$ and $V_{s}$ respectively, are related to the elastic Lam\'{e} parameters $\lambda$ and  $\mu$, through 
\begin{align}
\lambda + 2\mu & = \rho V_{p}^{2}, \label{eqn6} \\
 \mu & = \rho V_{s}^{2}. \label{eqn7}
\end{align}

\subsection*{Attenuation of seismic waves in viscoelastic media}
The Earth does not behave purely elastically, since waves attenuate---due to several dissipation mechanisms---as they propagate through the earth medium \citep{aki2002quantitative}. The attenuation of seismic energy due to the viscosity of earth rocks leads to a decrease in seismic-signal amplitude and phase dispersion of the recorded waveforms. This attenuation phenomenon can be modeled by viscoelastic mechanical models that usually contain weightless springs, which store strain energy, and dashpots, which dissipates energy \citep{carcione2014wave}. In this study, we adopt the Kelvin-Voigt model, which is a linear model and well justified in modeling the viscoelastic behavior of solid rocks in typical seismic band \citep{ba2014seismic,zhao20172}. Kelvin-Voigt viscoelasticity is introduced to build a viscoelastic medium based on an isotropic elastic material.

In the context of Kelvin-Voigt model, the dimensionless attenuation quality factor Q that characterizes energy dissipation in a material is given by $Q(\omega) = (\omega \tau)^{-1}$, $\tau$ being the relaxation time \citep{carcione2014wave}. A more detailed explanation of wave propagation in viscoelastic media and the implementation of the Kelvin-Voigt model of viscoelasticity in the frequency domain can be found in the literature \citep{carcione20043,ba2014seismic,carcione2014wave}.

\subsection*{Acoustic approximation of elastic wave propagation}
The acoustic wave equation can be derived as a special case of equation (\ref{eqn1}) by assuming the S-wave velocity $V_{s}$ is zero \citep[e.g.,][]{cance2015validity,monkola2016accuracy}. In such a case, only the diagonal elements of the stress tensor $\boldsymbol{\sigma }$ are non-zero and equal to the negative pressure $-p$. Hence, in acoustic media, equation (\ref{eqn1}) becomes
\begin{equation}
-\omega ^{2}\rho (\mathbf{x})\mathbf{u}(\mathbf{x},\omega ) = \nabla \cdot p(\mathbf{x}) + \mathbf{f}(\mathbf{x},\omega ).
\label{eqn003}
\end{equation} 
Substituting $\mu =0$ into (\ref{eqn4}) and (\ref{eqn6}) yield, respectively, the constitutive relations:
\begin{equation} \label{eq1111}
\begin{split}
\nabla \cdot \mathbf{u} & = -p/\lambda, \\
 V_{p} & = \sqrt{\lambda /\rho }. 
\end{split}
\end{equation}
Thus, taking the divergence of equation (\ref{eqn003}), assuming constant density, and substituting relations (\ref{eq1111}) leads to the Helmholtz acoustic wave equation, 
 \begin{equation}
- \nabla^{2}p(\mathbf{x}) - k^{2}(\mathbf{x})p(\mathbf{x})=f(\mathbf{x}),
\label{eqn8}
\end{equation}  
where $f = -\nabla \cdot\mathbf{f}$ is the dipole source term, and the wavenumber $k(\mathbf{x})$ of a lossless medium is related to $\omega$ by the usual formula 
\begin{equation}
k(\mathbf{x}) = \frac{\omega }{V_{p}(\mathbf{x})}.
\label{eqn111}
\end{equation}  

In general, the acoustic wave equation (\ref{eqn8}) can have the same solution as the elastic wave equation (\ref{eqn1}) only in the case for an infinite homogeneous isotropic medium, and assuming an explosive isotropic source that generates P-waves alone. However, modeling of elastic wave propagation in realistic earth should accommodate at least the free surface boundary condition and material interfaces. If there is a velocity contrast between layers, then reflections occur at each interface, producing S-wave reflections, breaking the assumption that the acoustic wave equation should not yield S-waves. Additionally, the acoustic approximation does not account for P-S (and S-P) mode-converted reflections, which implies that the amplitudes for the elastic P-waves are incorrectly modeled. 

In this study, we apply Bayesian-based acoustic Born inversion, that is acoustic inversion based on the Born approximation and the Bayesian framework, to the data calculated in the viscoelastic media. Hence, we neglect the elasticity and viscous effects in the Born waveform inversion because of computational efficiency. This can have dire consequences on the quality of the
recovered P-wave velocity ($V_{p}$) field of the subsurface due to the viscoelastic nature of the earth. But we adopt the Bayesian approximation error approach (discussed in Section \ref{BAE}), which takes into account the errors and uncertainties related to using Born (and acoustic) approximations to viscoelastic waves.  

\subsection{Born approximation} \label{Bornapp}
The first-order Born approximation is a single scattering approximation that is very attractive in the inversion of seismic data because it yields linear relations between parameters of interest and data. To derive this `single scattering' approach, we start by using the scalar Helmholtz equation (\ref{eqn8}) and decomposing the heterogeneous medium into a homogeneous background medium and the perturbations. Let us decompose the wave propagation velocity $V_{p}(\mathbf{x})$ into an average background velocity $V_{0}(\mathbf{x})$ and a perturbation $\chi (\mathbf{x})$ such that $k^{2}(\mathbf{x})$ in equation (\ref{eqn8}) can be expressed in terms of the constant $k_{0}$ for the background medium:
\begin{equation}
k^{2}(\mathbf{x}) = k_{0}^{2}[1+\chi (\mathbf{x})],
\label{eqn1132}
\end{equation}
where $k_{0} = \omega/V_{0}$. The perturbation $\chi (\mathbf{x})$ can be expressed as
\begin{equation}
\chi (\mathbf{x})=\left [ \frac{V_{0}^{2}}{V_{p}^{2}(\mathbf{x})}-1 \right ].
\label{eqn113}
\end{equation}

The Helmholtz equation for the background medium describes the propagation of the incident wavefield  $p_{0}(\mathbf{x})$ and is given by 
\begin{equation}
-\nabla^{2}p_{0}(\mathbf{x})-k_{0}^{2}p_{0}(\mathbf{x})=f(\mathbf{x}).
\label{eqn117}
\end{equation} 

The scattered wavefield $p_{s}(\mathbf{x})$ generated by the perturbations of the medium is given as follows:
\begin{equation}
p_{s}(\mathbf{x}) = -k_{0}^{2}\int_{D}\textsl{g}(\mathbf{x},\mathbf{x}{}')\chi (\mathbf{x}{}')\left [ p_{0}(\mathbf{x}{}')+ p_{s}(\mathbf{x}{}') \right ] d\mathbf{x}{}',
\label{eqn115}
\end{equation} 
where $D$ is the scattering domain where $\chi (\mathbf{x})$ is non-zero, and the Green's function $\textsl{g}(\mathbf{x},\mathbf{x}{}')$ is the wavefield at position $\mathbf{x}$ due to a point source at position $\mathbf{x}{}'$ in the background model.

For a homogeneous background with propagation speed $V_{0}$, the computation of $p_{0}$ becomes analytic. Substituting the approximation $p_{0} + p_{s}\approx p_{0}$ into the right-hand side of equation (\ref{eqn115}) yields the \textbf{Born approximation}. The validity of this approximation has been explored in the literature \citep[e.g.,][]{habashy1993beyond,parisi2014validity,chen2015full}. As a weak scattering approximation, the Born approximation does not account for multiple scattering effects, it is not suitable for high scattering contrasts, and it is only valid in the low frequency regime with respect to the scattering domain. In this study, the BAE approach will be used to overcome these limitations associated with the Born approximation. 

\section{The seismic forward problem} \label{Section3}
Here, we simulate data assuming that all input parameters are known, a procedure often referred to as the forward problem. Hence, we deal with the solution of the wave equation (with specified initial and boundary conditions), given the velocity field as well as the mathematical representation of the source and the source--receiver configuration.  

We use COMSOL Multiphysics$^{\mbox{\scriptsize\textregistered}}$ (Finite Element solver) to numerically solve the viscoelastic wave equation in the frequency domain for a given problem setup. We take advantage of the Structural Mechanics Module that provides modeling tools for linear elastic and viscoelastic material models. To minimize undesirable reflections from the boundaries of the computation domain, the following boundary conditions are applied: a free surface at the top of the model, and absorbing boundary conditions to the sides and bottom of the model. The absorbing boundary condition used at the computational edges is achieved with Perfectly-Matched Layers (PML). 

To solve the linearized acoustic wave equation, which is used in the estimation stage, the forward problem involves solving $p_{s}(\mathbf{x})$ from equation (\ref{eqn115}). First, we discretize the model geometry by $N =N_{x} \times N_{z}$ grid cells of equal area $\Delta A_{n},\ n\in \left \{1,\ldots,N \right\}$, where $N_{x}$ and $N_{z}$ are numbers of grid cells in $x$ and $z$ directions, respectively. The chosen mesh size should be much smaller than the smallest wavelength of the seismic waves to obtain suitable numerical accuracy. We assume that the scattered wavefield data for a discrete set of $N_{f}$ frequencies $N_{f}, \ f\in \left \{ 1,\ldots,N_{f} \right \}$ is generated from $N_{s}$ sources $N_{s}, \ s\in \left \{1,\ldots,N_{s} \right \}$ and recorded by $N_{r}$ receivers $N_{r},\  r\in \left \{ 1,\ldots,N_{r} \right \}$. 
After discretization \citep{muhumuza2018seismic}, we obtain the following equation for the scattered wavefield in equation (\ref{eqn115}):
\begin{equation}
p_{s}^{(sf)} = \sum_{n=1}^{N}G_{rn}^{(sf)}\chi _{n},
\label{eqn118}
\end{equation} 
where $\chi _{n}$ is the $n$th component of the parameter vector $\mathbf{m} = (\chi_{1},\ldots,\chi_{N})$ representing squared slowness perturbation values in equation (\ref{eqn113}) for each $n$th grid cell, and 
\begin{equation}
G_{rn}^{(sf)} = \Delta A_{n} k_{0}^{2}\textsl{g}(\mathbf{x}_{r},\mathbf{x}_{s})  p_{0}(\mathbf{x}_{r},\mathbf{x}_{s}).
\label{eqn119}
\end{equation} 
If we now reduce the three indices $f, s, r$ to one index $\beta$, i.e., $f, s, r \rightarrow  \beta = 1,...,N_{d}=N_{f}\times N_{s}\times N_{r}$, then we can write equation (\ref{eqn118}) in matrix notation:
\begin{equation}
\mathbf{d}= \mathbf{G} \mathbf{m},
\label{eqn120}
\end{equation} 
where $\mathbf{d}$ is the scattered data vector of length $N_{d}$ containing all the different frequency components of the scattered wavefield for all sources and all receivers. Therefore, the Born approximation linearizes equation (\ref{eqn115}) to a linear problem (\ref{eqn120}), which is readily soluble using the Bayesian approach to linear inverse problems. 

In our implementation of the Born approximation, we use the method of images \citep{1999fuacbook} to implement the free surface (Dirichlet boundary condition), but absorbing boundary layers on the sides of the model are not evoked. The resulting modeling errors caused by spurious reflections from the lateral boundaries of the model will be accounted for using the BAE approach. 

\section*{The seismic inverse problem} \label{Section4}

In this paper, we aim to  reconstruct the $V_{p}$ field for viscoelastic media by employing the acoustic Born approximation, which reduces computational cost. As described in section \ref{Bornapp}, the Born approximation  relies on linearizing the problem with respect to the squared slowness perturbation $\chi$. Therefore, instead of estimating $V_{p}$ directly, the linearized inverse problem seeks to estimate the parameter vector $\mathbf{m} = (\chi_{1},\ldots,\chi_{N})$ from the seismic data $\mathbf{d}$. The estimates are then converted back to $V_{p}$ for visualization purposes.

The inverse problem can be solved in the Bayesian framework 
\citep{tarantola2005inverse,kaipio2006statistical}, where all unknowns and data are modeled as random variables. The solution is based on combining information coming from the observed data $\mathbf{d}$ and the model parameters of interest $\mathbf{m}$ of the examined medium, offering a framework for uncertainty quantification. There are two main tasks here: constructing the likelihood model (conditional density of the data given the parameters) $\pi(\mathbf{d}|\mathbf{m})$, and determining the prior density (distribution of the parameters in the absence of any data) $\pi(\mathbf{m})$. The posterior probability density $\pi(\mathbf{m}|\mathbf{d})$ is given by the Bayes' theorem,
\begin{equation}
\pi(\mathbf{m}|\mathbf{d}) = \frac{\pi(\mathbf{d}|\mathbf{m})\pi(\mathbf{m})}{\int \pi(\mathbf{d}|\mathbf{m})\pi(\mathbf{m})d\mathbf{m}} \propto \pi(\mathbf{d}|\mathbf{m})\pi(\mathbf{m}),
\label{eqn10}
\end{equation} 
where the denominator $\int \pi(\mathbf{d}|\mathbf{m})\pi(\mathbf{m})d\mathbf{m}$ is a normalizing constant, which can usually be ignored. 
In principle, the posterior can be (mathematically) considered as the solution of the problem, but practical solutions typically seeks for point estimates that are computed based on the posterior distribution.

The statistical extension of (\ref{eqn120}) can be written as 
\begin{equation}\
\mathbf{d}= \mathbf{G} \mathbf{m}+e,
\label{eqn120_noise}
\end{equation} 
where $e$ is a random variable representing observation noise. If we assume that the term $e$ is Gaussian distributed with zero mean and independent of $\mathbf{m}$, the likelihood model can be written as 
\begin{equation}
\pi(\mathbf{d|\mathbf{m}}) \propto \textup{exp}\left \{ -\frac{1}{2}\left \| L_e(\mathbf{d}-G\mathbf{m}) \right \|^{2} \right\},
\label{eqn15_2}
\end{equation} 
where $L_e$ is computed from the factorization of the inverse of the noise covariance matrix
$C_e^{-1} = L_e^{T}L_e$. Factorization can be done, for example, with Cholesky decomposition.

The above model (\ref{eqn15_2}), however, does not include possible modelling errors. Because we employ a highly approximative acoustic Born model in the inversion procedure for viscoelastic media, the related approximation/modelling errors must be treated accordingly for reliable estimates to be obtained. Therefore, here we use the BAE approach that allows us to incorporate all the approximation and modeling errors resulting from the approximative model into the posterior distribution. 

\subsection*{Bayesian approximation error approach}\label{BAE}

The Bayesian approximation error approach relies on two computational forward models: an accurate model, which is accurate in the sense that its modelling errors are negligible compared to other errors such as observation noise, and an approximate computationally cheap model. In this work, the elastic-viscous model $\overbar{G}$ is taken as the accurate model and the computationally cheap model is given by the acoustic Born approximation $G$.

We can write the observation model for the BAE approach in the form
\begin{equation} \label{eqn11}
\mathbf{d} = \overbar{G}(\overbar{\mathbf{m}}) + 	e = G\mathbf{m} + \varepsilon + e, 
\end{equation} 
where $\overbar{\mathbf{m}}$ is the vector of accurate model parameters. Here $\varepsilon = \overbar{G}(\overbar{\mathbf{m}})-G\mathbf{m}$ represents the approximation error term, which is the
discrepancy between predictions of the scattered wavefield (for a known scattering medium and incident field) when using the accurate elastic-viscous model $\overbar{G}(\overbar{\mathbf{m}})$ and the approximate acoustic Born model $G\mathbf{m}$. In our implementation, the accurate nonlinear forward modeling $\overbar{G}(\overbar{\mathbf{m}})$ is specified by velocities $V_{p}$, $V_{s}$ and quality factor $Q$, while the linear forward modelling $G\mathbf{m}$ is specified only by $V_{p}$. 

Hence, the main idea behind the BAE approach is to replace the computationally demanding accurate mapping $\overbar{G}$ by $G$ that is less accurate but computationally feasible by taking into account induced errors through the error term $\varepsilon$. Although we apply BAE here to seismic imaging, the method is more general and could be applied for any two models that differ but have the same approximate solution \citep[e.g.,][]{nissinen2007bayesian,koponen2014bayesian}. In principle, we require that the absolute value of the difference $\left |\overbar{G}(\overbar{\mathbf{m}})- G\mathbf{m} \right | \ll \left | G\mathbf{m} \right |$; and that, the approximation and model errors result only from the forward model. 

We denote by $\theta\sim \mathcal{N}(\theta_{*},C_{\theta})$ the multivariate joint normal distribution with mean $\theta_{*} = \mathbb{E}(\theta)$ and covariance $C_{\theta} = \textup{Cov}(\theta)$. In the BAE approach, we define an additive error term as $\nu = \varepsilon+e$ and approximate both $e$ and the conditional density $\nu$ given the parameter of interest $\mathbf{m}$ as Gaussian distributions, i.e., $e\sim \mathcal{N}(e_{*},C_{e})$ and $\nu |\mathbf{m}\sim \mathcal{N}(\nu _{*|\mathbf{m}},C_\mathbf{\nu |m})$. Hence,
\begin{equation}
\nu_{*|\mathbf{m}} = \nu _{*}+ C_{\varepsilon \mathbf{m}} C_{\mathbf{m}}^{-1}(\mathbf{m}-\mathbf{m}_{*}), 
\label{eqn12}
\end{equation} 
\begin{equation}
C_{\nu |\mathbf{m}} = C _{e}+ C_{\varepsilon}-C_{\varepsilon\mathbf{m}} C_{\mathbf{m}}^{-1}C_{\mathbf{m}\varepsilon },
\label{eqn13}
\end{equation} 
where $C_{\varepsilon\mathbf{m}}=C^{'}_{\mathbf{m}\varepsilon}$ is the cross-covariance matrix of $\varepsilon$ and $\mathbf{m}$, and $C_{\mathbf{m}}^{-1}$ is the prior model inverse covariance.

On the assumption that the measurement errors are mutually independent with the model parameters, the observational model (\ref{eqn11}) leads to a likelihood model
\begin{equation}
\mathbf{d}|\mathbf{m}\sim \mathcal{N}(\mathbf{d}-G\mathbf{m}-\nu _{*|\mathbf{m}},C_{\nu |\mathbf{m}}).
\label{eqn14}
\end{equation} 
To further simplify the analysis, we adopt an approximation called enhanced error model \citep{kaipio2006statistical,kaipio2007statistical}, which is obtained by setting $C_{\varepsilon\mathbf{m}}$ to zero. Thus, $\nu _{*|\mathbf{m}} \approx \nu _{*}=e_{*}+\varepsilon _{*}$ and $C_{\nu |\mathbf{m}} \approx C_{\nu }=C_{e}+C_{\varepsilon }$. This further approximation is commonly used to stabilize the numerical approximation of $C_{\nu |\mathbf{m}}$. 

In order to compute the statistics of the approximation errors $\varepsilon$ using the accurate and approximate forward modeling procedures, we generate samples of $V_{p}$ and $V_{s}$ fields by employing a level-set-based model with Gaussian process as explained in section \ref{samples}. We then compute the scattered fields for acoustic Born model $G\mathbf{m}$ and elastic-viscous model $\overbar{G}(\overbar{\mathbf{m}})$. The respective approximation errors of $H$ generated samples,
$\varepsilon^{(a)}=\overbar{G}(\overbar{\mathbf{m}^{(a)}})-G\mathbf{m}^{(a)},\  a\in \left \{ 1,\ldots,H \right \}$ are then obtained and used to compute the sample mean $\varepsilon _{*}$ and sample covariance $C_{\varepsilon}$.

\subsection*{Prior models for inversion}

In this paper, we consider two choices for the prior: the Gaussian anisotropic smoothness prior \citep{williams2006gaussian} and the non-Gaussian Cauchy prior \citep{markkanen2019cauchy,mendoza2019statistical}. The Gaussian assumption about the prior distribution of the acoustic/elastic parameters coupled with a Gaussian likelihood is commonly used for inversion of seismic data because it yields an analytical expression for the posterior distribution. However, subsurface velocity structure usually contain sharp interfaces that can be difficult to reconstruct using Gaussian priors. The use of the edge-preserving Cauchy prior could potentially overcome this challenge  \citep{markkanen2019cauchy}. 

First, we consider the Gaussian prior: 
we assume that $\mathbf{m} \sim \mathcal{N}(\mathbf{m}_{*},C_{\mathbf{m}})$. Here, the prior mean $\mathbf{m}_{*}$ and covariance $C_{\mathbf{m}}$ are chosen based on experience and prior knowledge about the parameters of interest. In this Gaussian case, the posterior distribution of the model vector  $\mathbf{m}$ conditioned by the seismic data  $\mathbf{d}$ can be  written as 
\begin{equation}
\pi(\mathbf{m}|\mathbf{d}) \propto \textup{exp}\left \{ -\frac{1}{2}\left \| L_{v|\mathbf{m}}(\mathbf{d}-G\mathbf{m}-\nu _{*|\mathbf{m}}) \right \|^{2}-\frac{1}{2}\left \| L_{\mathbf{m}}(\mathbf{m}-\mathbf{m}_{*}) \right \|^{2} \right \},
\label{eqn15}
\end{equation} 
where $L_{v|\mathbf{m}}$ and $L_{\mathbf{m}}$ are the Cholesky decompositions of the inverse covariance matrices, i.e., $C_{\nu |\mathbf{m}}^{-1} = L_{\nu |\mathbf{m}}^{T}L_{\nu |\mathbf{m}} 
$ and $C_{\mathbf{m}}^{-1} = L_{\mathbf{m}}^{T}L_{\mathbf{m}}$, and $\nu _{*|\mathbf{m}} =\nu _{*|\mathbf{m}}(\mathbf{m})$. 
As the observation model $G\mathbf{m}$ is linear, the posterior is Gaussian with the mean corresponding to the maximum point of the distribution:
\begin{equation}
\mathbf{m}_{\textup{CM}}^{BAE} =\underset{\mathbf{m}}{\textup{arg min}}\left \| L_{v|\mathbf{m}}(\mathbf{d}-G\mathbf{m}-\nu _{*|\mathbf{m}}) \right \|^{2}+\left \| L_{\mathbf{m}}(\mathbf{m}-\mathbf{m}_{*}) \right \|^{2}.
\label{eqn16_gen}
\end{equation}
The mean of the posterior is often called as the conditional mean (CM) estimate.
The optimization problem has a practical closed form solution (c.f. linear LS-estimation).

The covariance of the posterior reflecting the uncertainty of the estimate can be calculated as \citep[e.g., see][]{damien2013bayesian}
\begin{equation}
C_{\mathbf{m}|\mathbf{d}}^{\textup{BAE}} = \left (G^{T}C_{\nu|\mathbf{m} }^{-1}G+ C_{\mathbf{m}}^{-1} \right )^{-1}.
\label{eqn17_gen}
\end{equation}
Without BAE, the conventional CM estimate $\mathbf{m}_{\textup{CM}}^{CEM}$ and posterior covariance $C_{\mathbf{m}|\mathbf{d}}^{\textup{CEM}}$ can be computed similarly as in (\ref{eqn16_gen}) and (\ref{eqn17_gen}) but ignoring the BAE term $\nu _{*|\mathbf{m}}$ (i.e. $L_{v|\mathbf{m}}=L_{e}$, $\nu _{*|\mathbf{m}}=e_{*}$, and $C_{\nu|\mathbf{m} }^{-1}=C_{e}^{-1}$)

Second, we consider the non-Gaussian Cauchy prior:
The edge-preserving Cauchy prior can be expressed as a product
\begin{equation}
    \pi(\mathbf{m}) = \prod_{j=1}^{N_{x}} \prod_{j'=1}^{N_{z}} \frac{\lambda h}{(\lambda h)^2+(\mathbf{m}_{j,j'}-\mathbf{m}_{j-1,j'})^2}
    \frac{\lambda h'}{(\lambda h')^2+(\mathbf{m}_{j,j'}-\mathbf{m}_{j,j'-1})^2}, 
\end{equation}
where $(j,j^{'})\in \mathbb{I}^{2}\subset \mathbb{Z}^{2}$, $\lambda$ is the regularization parameter, and $h,h'>0$ are discretization steps in the $x$- and $z$-directions.
For more details on the numerical implementation of these priors, see \cite{markkanen2019cauchy,mendoza2019statistical}.
In a similar setting, for theoretical posterior consistency analysis with respect to mesh refinement  for Cauchy and more general L\'evy alpha-stable sheet priors, see \citet{chada2019}.
The typical alternatives for edge-preserving inversion are total variation \citep[see e.g.][]{Rubin1992} and Besov priors \citep{Lassas2009}.
However,  TV priors are not consistent under mesh refinement \citep{Lassas2004} and the Besov priors rely on wavelets  that are difficult to implement with pixel-based approximations.
Alternatively we could also use level set methods \citep{chada2018parameterizations,dunlop2017hierarchical}, but they typically require to predefine the number of the level sets (subdomains) prior to the prediction which we prefer to avoid.

For the non-Gaussian Cauchy prior, it is not possible to obtain the posterior distributions analytically. 
In this work, we use Markov chain Monte Carlo (MCMC) sampling to explore the posterior distribution and obtain CM estimates with uncertainty quantification. 
We follow the practice in  \citet{markkanen2019cauchy}, and use Metropolis-within-Gibbs for drawing samples form the posterior.

\subsection{Interval estimation}
\label{sec:interval_estimation}

We can summarize the marginal posterior distributions by tabulating the credible
intervals of the estimated parameter $\mathbf{m}$ based on standard deviation. For example, if $\mathbf{m}_{i}$ is the $i$th element of the estimated parameter vector $\mathbf{m}$, then the $95\%$  credible interval for $\mathbf{m}_{i}$ is computed as 
\begin{equation}
\left [ \mathbf{m}_{*|\mathbf{d},i}-1.96 \sigma_{*|\mathbf{d},i},\, \mathbf{m}_{*|\mathbf{d},i}+ 1.96\sigma_{*|\mathbf{d},i} \, \right ],
\label{eqn20}
\end{equation}
where $\mathbf{m}_{*|\mathbf{d},i}$ represents the $i$th component of $\mathbf{m}_{i}|\mathbf{d}$ and $\sigma_{*|\mathbf{d},i}$ is the standard deviation of the estimate. 
In the case of a Gaussian prior, $\sigma_{*|\mathbf{d},i}$ 
can be obtained from the diagonal of the covariance estimate $C_{\mathbf{m}|\mathbf{d}}^{\textup{BAE}}$, and, in the case of a Cauchy prior, it can be estimated as the standard deviation of the MCMC samples.

\section*{NUMERICAL EXPERIMENTS} \label{Section5}

\subsection{Setup of test cases}
\begin{figure}[H]
\begin{center}
\includegraphics[width=0.8\textwidth]{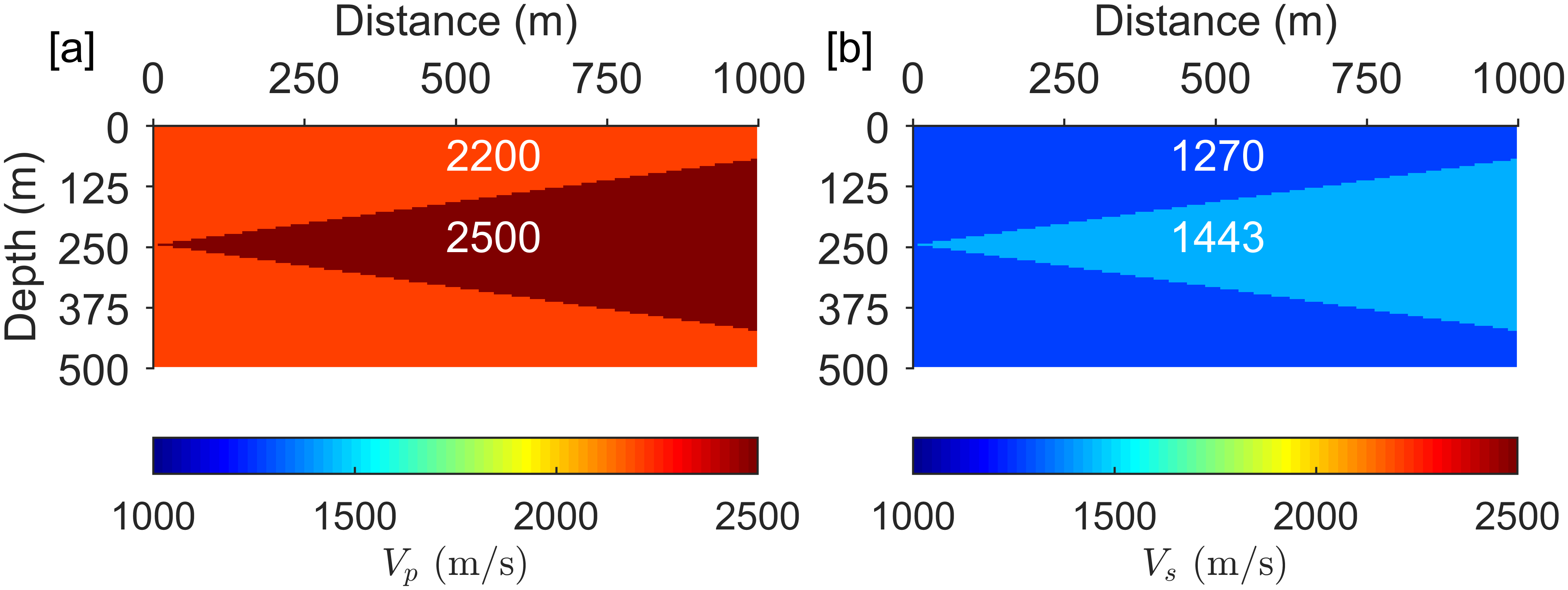}
\end{center}
\caption{The true velocity field of medium A: (a) P- and (b) S-wave velocities.}
  \label{modelA}
\end{figure}

In order to demonstrate our inversion framework, we consider two 2D experiments of  viscoelastic isotropic media characterized by varying degrees of complexity. The first (medium A) is a single wedge whose $V_{p}$ field is constructed (in Figure~\ref{modelA}a)
with a high-velocity wedge embedded between two low-velocity beds. The S-wave velocity field $V_{s}$ is calculated by the $V_{s} \approx V_{p}/\sqrt{3}$ criteria. The density of the medium is chosen to be constant at 2000 $\textup{kg}/\textup{m}^{3}$ which is typical for sedimentary rocks. 

\begin{figure}[H]
\begin{center}
\includegraphics[width=0.8\textwidth]{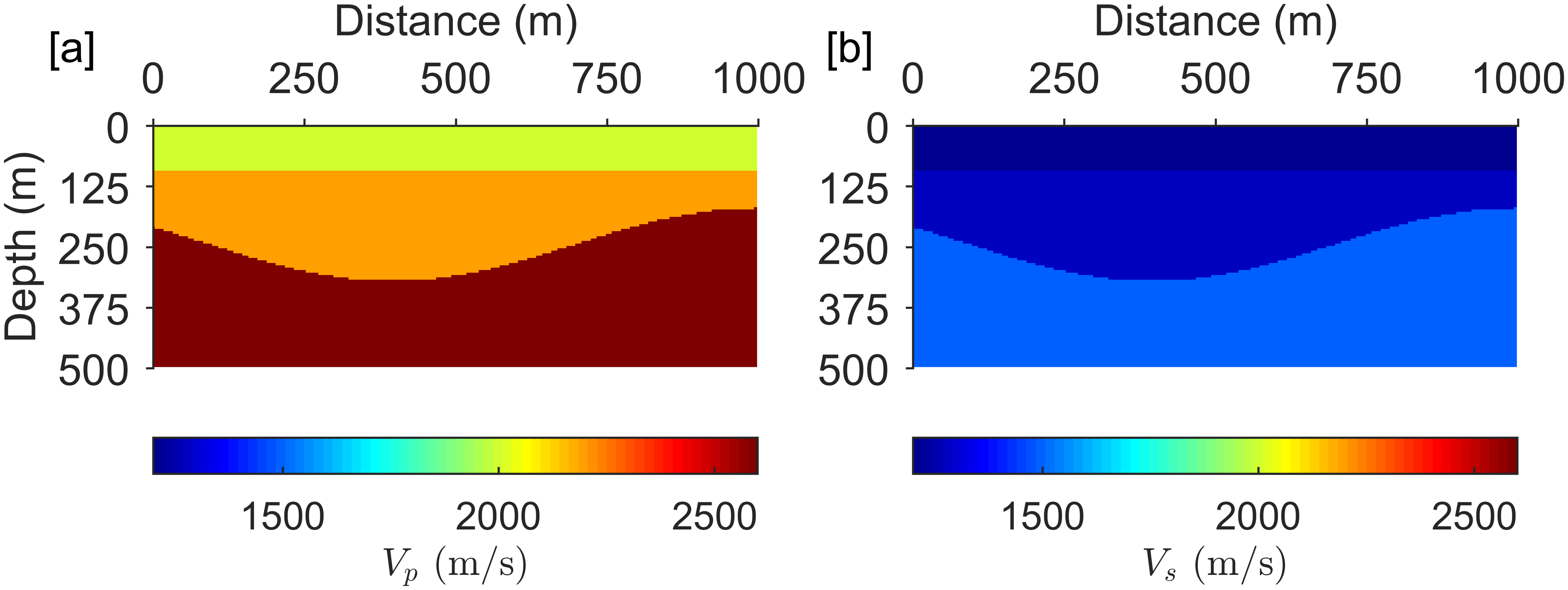}
\end{center}
\caption{The true velocity field of medium B: (a) P- and (b) S-wave velocities.}
  \label{SEGmodel1}
\end{figure}

The second (medium B) is a three-layered medium with a syncline interface shown in Figure~\ref{SEGmodel1}. The model dimensions are the same as in model A. The S-wave velocities are built from the P-wave velocity model so that Poisson's ratio is constant at 0.25, resulting
in an S-wave velocity model in Figure~\ref{SEGmodel1}b. The density of all the layers of the medium is also chosen to be constant at 2000 $\textup{kg}/\textup{m}^{3}$. 

\subsection*{Simulation of measurement data}\label{simulation}

We use COMSOL Multiphysics software to generate synthetic seismic data (vertical component of particle acceleration) for 2D viscoelastic isotropic media A and B. We apply absorbing boundary conditions to the lateral domain and bottom boundaries---achieved with the use of PML. A free-surface condition is used on the top boundary. For each medium, we simulate data for 8 frequencies used in inversion (1, 3, 5, 7.5, 10, 12, 15, and 18 Hz), with a 7.5 Hz Ricker wavelet source. There are 34 vertical-force sources (dipole sources) evenly distributed from $x = 7.5$ m to $x = 997.5$ m buried by 2.5 m. The wavefield generated by ``buried'' vertical-force sources is recorded by 100 evenly spaced receivers deployed at 5 m depth. We used a constant 2,200 m/s as background velocity. To obtain the scattered wavefield field $\mathbf{d}$, the incident wavefield is simulated and subtracted from the total wavefield.

For the simulated viscoelastic data, we have set the quality factor equal to $Q = 50$ for both P and S waves. To avoid both the inverse crime \citep{kaipio2006statistical} and numerical dispersion of the P- or S-waves, we used a grid of size 5 m with 6 elements per shortest wavelength. Second-order shape functions are used to give the best trade-off between model size and accuracy. 

As the measurement noise error model, we add zero mean white noise to the waveform data $\mathbf{d}$ in the frequency domain, with the noise covariance matrix given by $\boldsymbol{C}_e=\delta^2_e\boldsymbol{I}$ , and $\delta_e=(\max(\mathbf{d})-\min(\mathbf{d}))\times4/100$. This implies that the noise level is $4\%$  of the range of the noiseless data. 

\begin{figure}[H]
\begin{center}
\includegraphics[width=0.8\textwidth]{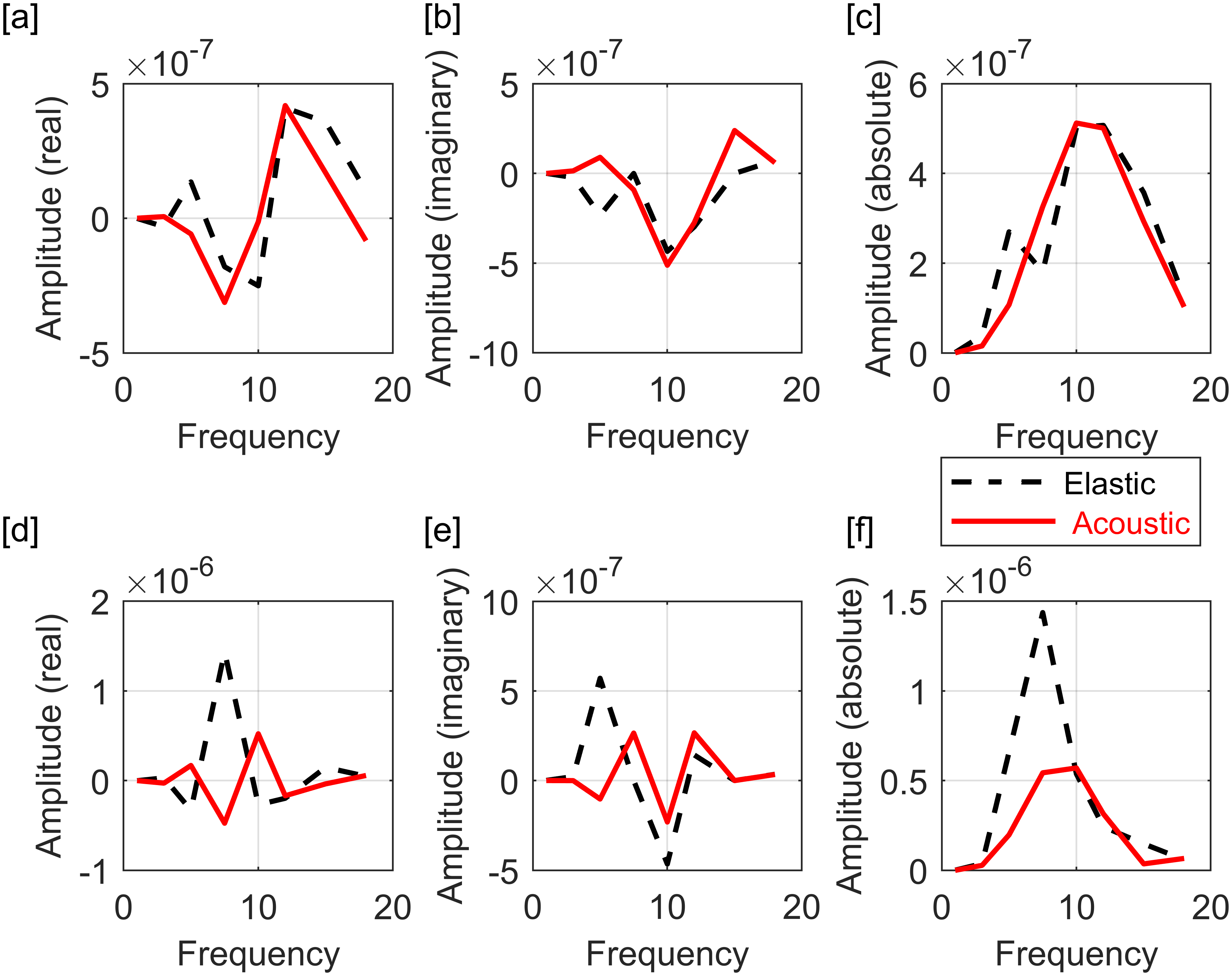}
\end{center}
\caption{Medium A: Comparison of acoustic (red) and viscoelastic (black) seismic traces in the frequency domain for offsets 112.5 m (a -- c) and 642.5 m (d -- f), respectively. The source point is at a distance of 67.5 m.} 
  \label{Compare}
\end{figure}

A comparison of acoustic and viscoelastic noiseless data computed from model A for a given source location and two offset positions is shown in Figure~\ref{Compare}. At near offsets and lower frequency, we observe a relatively better match in phases and amplitudes. However, at large offsets and higher frequencies, there are significant differences between viscoelastic and acoustic Born approximate data. This difference is due to both elastic effects and dispersion in the viscoelastic medium. The acoustic Born modeling does not therefore match the viscoelastic modeling and wrongly predicts the amplitudes and phases of the seismic data. 

\subsection*{Computation of approximation error statistics}\label{samples}

The BAE approach involves drawing random realizations (samples) of the velocity field and solving the forward problem for the draws with two models: accurate and approximate. Here, we utilized an accurate elastic-viscous model $\overbar{G}(\overbar{\mathbf{m}})$ based on the finite element solver, and an approximate acoustic model $G\mathbf{m}$ based on the Born approximation. The modeling parameters and source-receiver geometry remained as in the case for generating the synthetic measurement data in section \ref{simulation}. To relate the viscoelastic data with acoustic Born data, we carried out all of our simulations using dipole sources, and the vertical component of particle acceleration derives from pressure gradients.  

\begin{figure}[H]
\begin{center}
\includegraphics[width=0.8\textwidth]{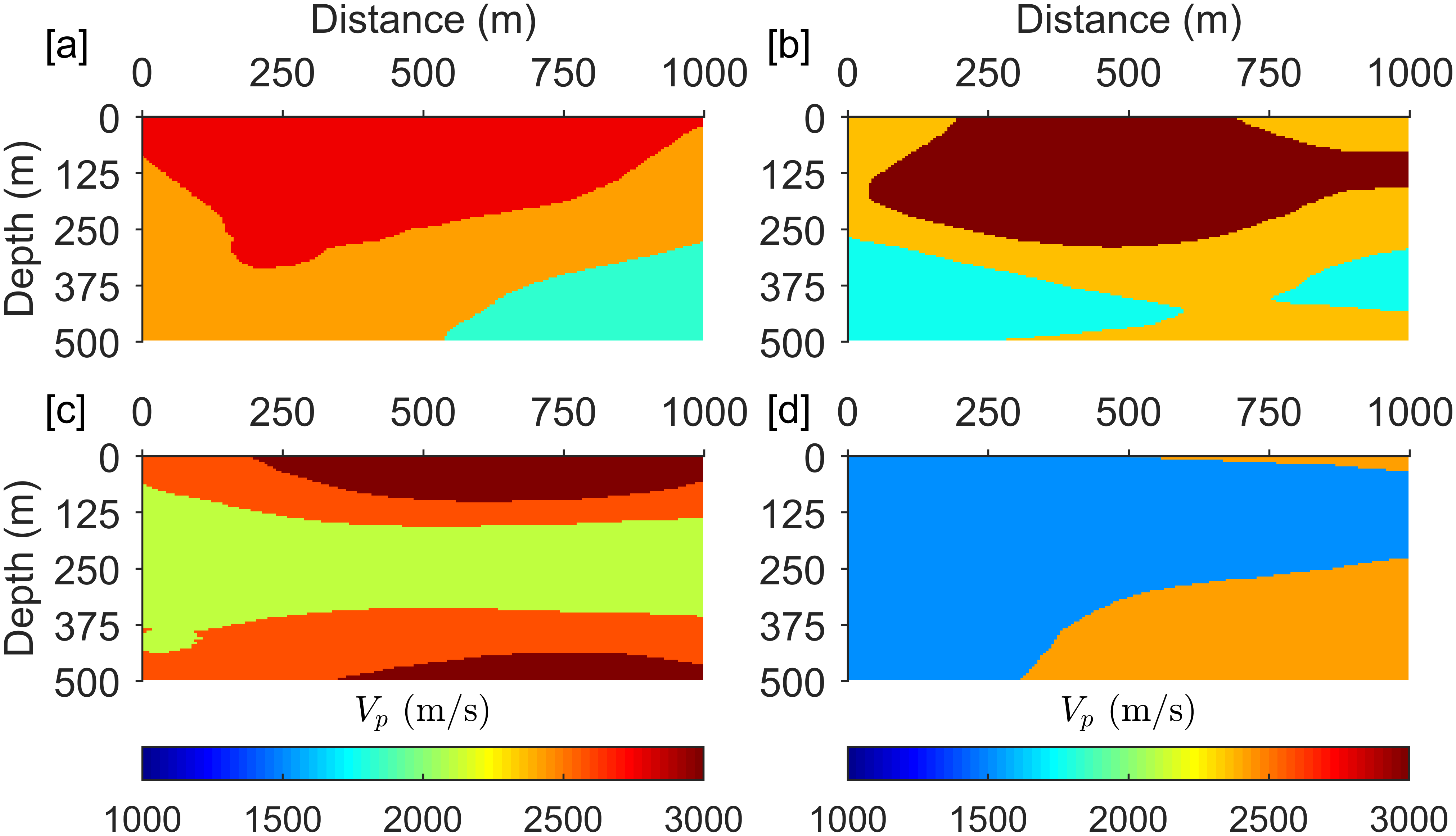}
\end{center}
\caption{Four random realizations of the P-wave velocity field. The anisotropic Gaussian process is used as the level set function to generate the samples.}
  \label{simplemodel1}
\end{figure}

We generated 5000 samples for BAE approach using a level set approach \citep{osher2001level} with Gaussian process as described in \cite{dunlop2017hierarchical} and \cite{chada2018parameterizations}. The procedure is described as follows: The velocity field $V_{p}$ is modelled via a level set function $\phi (\mathbf{x}),\mathbf{x}\in\mathbf{\Omega }$, where the boundary of the domain $\mathbf{\Omega }$ is represented as the zero level contour of $\phi (\mathbf{x})$. The level set function $\phi (\mathbf{x})$ is chosen to be a Gaussian random field \citep{williams2006gaussian} such that the smoothness in $\phi (\mathbf{x})$ translates into the interfaces given by the level sets $ \left \{ \mathbf{x}\mid\phi (\mathbf{x})\geqslant 0 \right \}$. The level-set parameters are selected to randomly produce one to three regions with different $V_{p}$ values, partitioning the domain $\mathbf{\Omega }$ into a number of sub-domains (Figure~\ref{simplemodel1}). The P-wave velocities in each sub-domain randomly ranged from 1200--3200 m/s. The S-wave velocities corresponding to each sub-domain are generated from the P-wave velocities such that the $V_{p}/V_{s}$ ratio varies randomly between 1.5 and 2.0, which are typical values for rocks in the crust. In our simulation of samples for the accurate elastic-viscous model, the values of the quality factor $Q$ randomly varied between 50 and 200. The smaller the quality factor the stronger the attenuation effect, and the higher the viscoelasticity as well.

It should be noted that the generation of the samples can be a computationally expensive task, but it can be carried out offline and has to be performed only once. When the BAE statistics are precomputed, then they are used for all BAE reconstructions presented in section \ref{Section5.4}.

\subsection*{Inversion results and discussion}\label{Section5.4}
To validate the feasibility of the BAE approach to partially recover from the errors induced by acoustic Born inversion in viscoelastic media, we apply our methodology to viscoelastic data generated from medium A and B. We performed the acoustic Born inversion for 8 frequencies ranging from 1 to 18 Hz, using an inversion grid size of $10$ m $\times 10$ m. 

\begin{figure}[H]
\begin{center}
\includegraphics[width=0.8\textwidth]{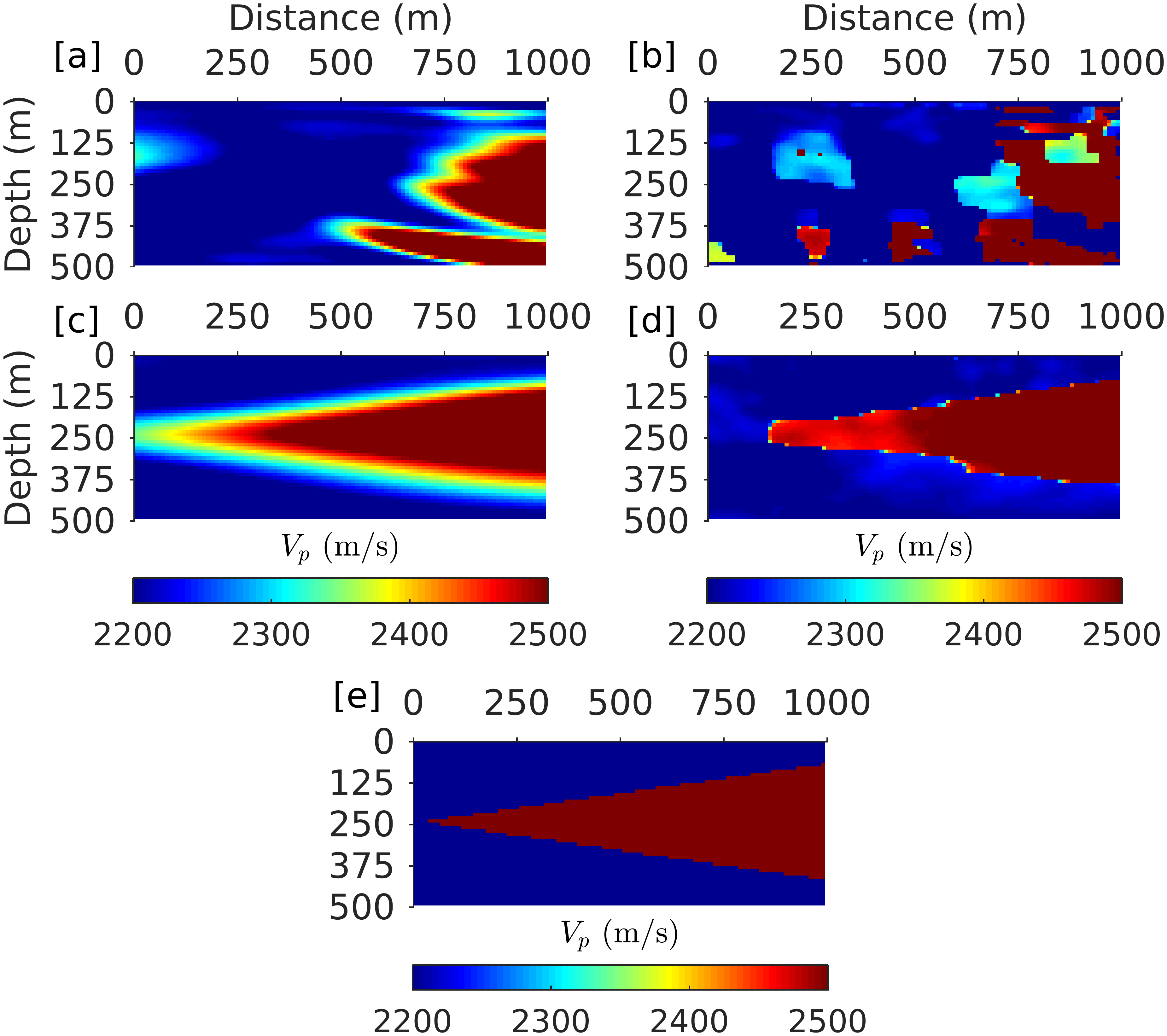}
\end{center}
\caption{Inversion results for medium A: Estimated velocity field corresponding to (a): CM estimate using the Gaussian prior without BAE (b): CM estimate using the Cauchy prior without BAE (c): CM estimate using the Gaussian prior with BAE (d): CM estimate using Cauchy prior with BAE and (e): the true P-wave velocity field.}
  \label{INVmodel}
\end{figure}

\begin{figure}[H]
\begin{center}
\includegraphics[width=0.8\textwidth]{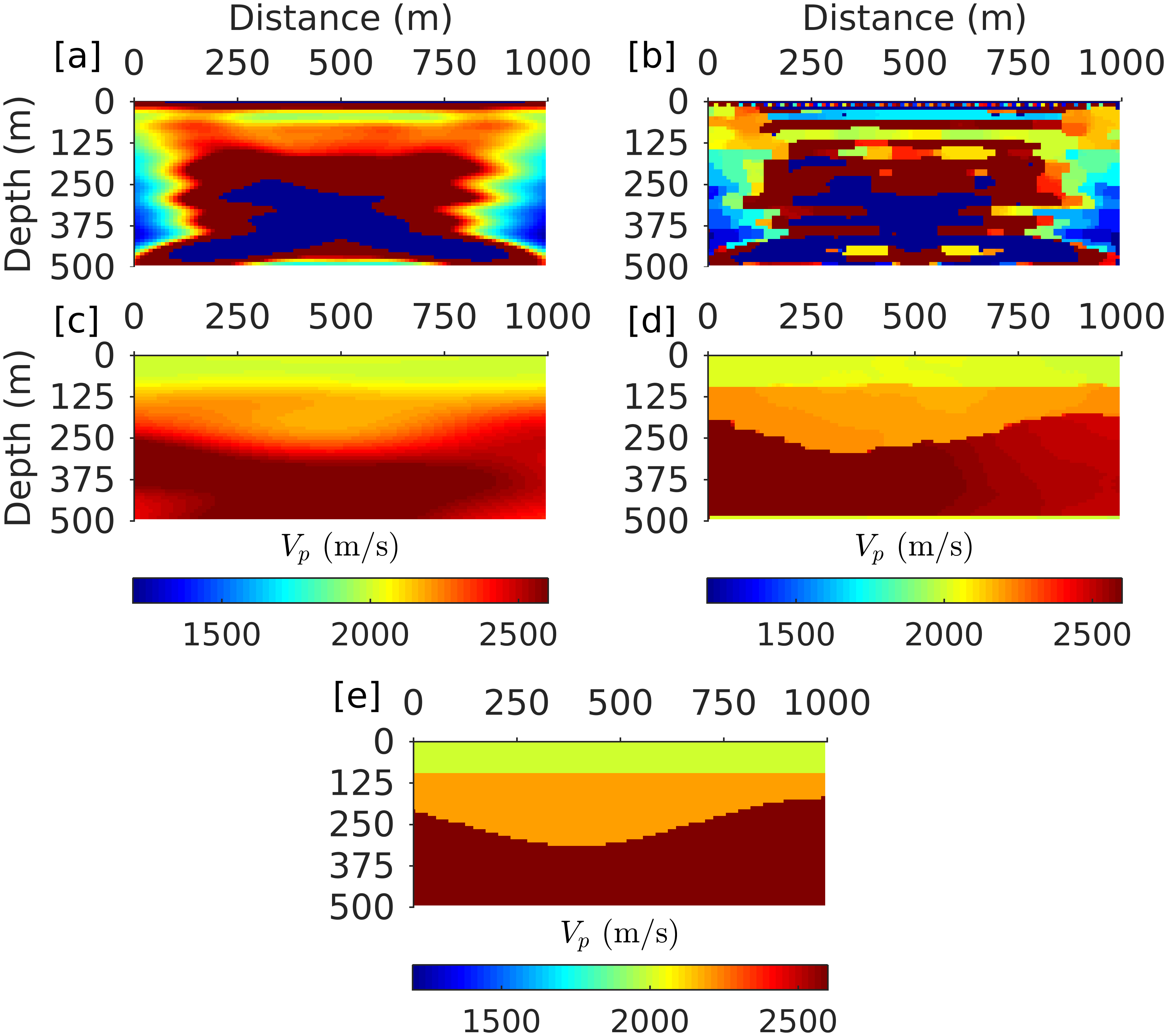}
\end{center}
\caption{Inversion results for medium B: Estimated velocity field corresponding to (a): CM estimate using the Gaussian prior without BAE (b): CM estimate using the Cauchy prior without BAE (c): CM estimate using the Gaussian prior with BAE (d): CM estimate using Cauchy prior with BAE and (e): the true P-wave velocity field.}

  \label{SEGINVmodel}
\end{figure}

We estimate the parameter vector $\mathbf{m}$ that contains perturbations in the squared slownesses $\chi$ from the data vector $\mathbf{d}$ using a simultaneous multi-frequency inversion approach. However, since seismic imaging often deals with velocities, we compute $V_{p}$ by equation (\ref{eqn113}) for the purpose of visualization. Using the two priors, the velocity fields corresponding to the CM estimates computed without and with the BAE approach are shown in Figures~\ref{INVmodel} and \ref{SEGINVmodel}. 

\begin{figure}[H]
\begin{center}
\includegraphics[width=0.8\textwidth]{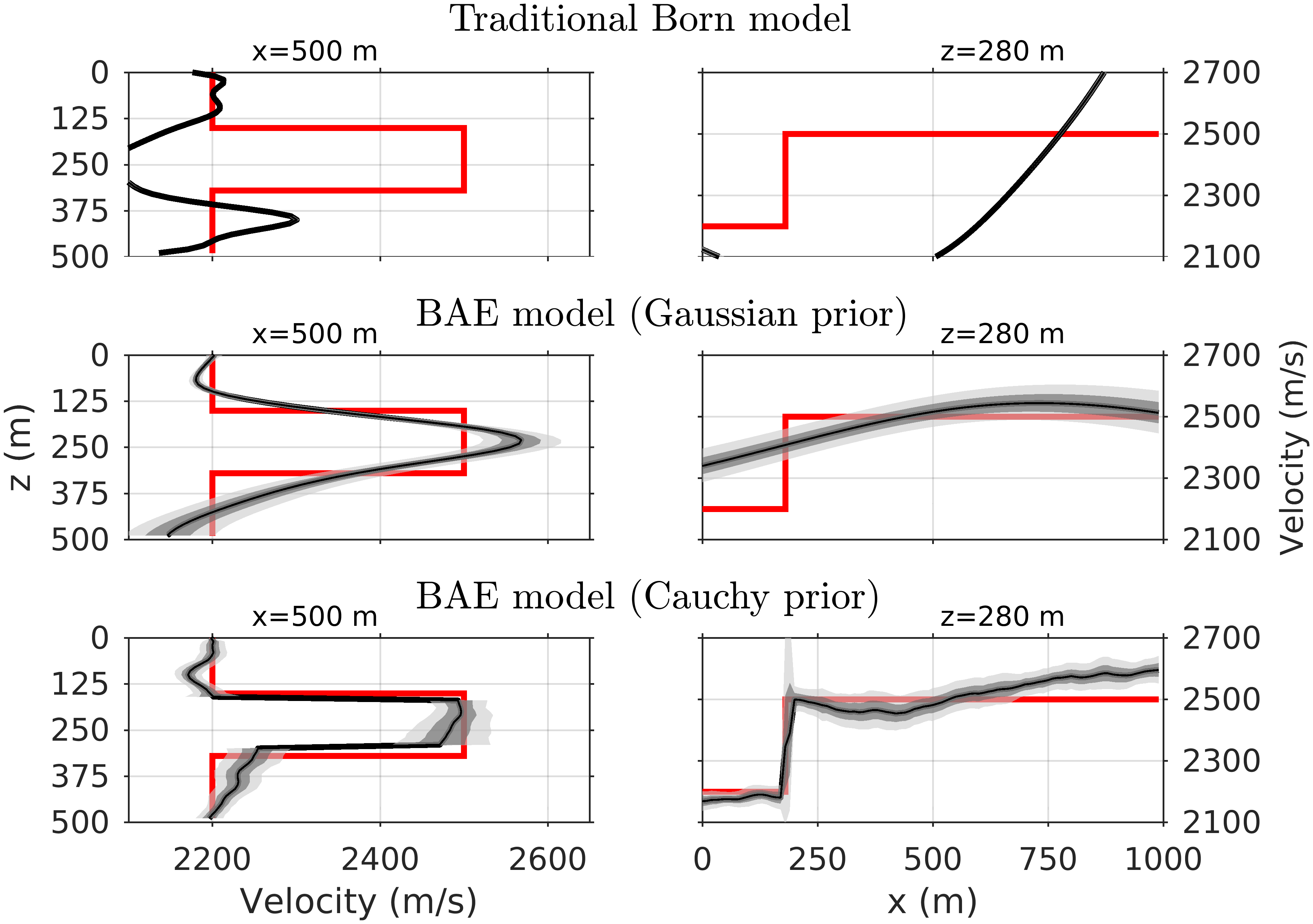}
\end{center}
\caption{Vertical and horizontal profiles of estimated P-wave velocity field for medium A along $x=500$ m (left panel) and $z=280$ m (right panel), respectively: Red lines show the true velocity profile while black lines show the profile corresponding to the CM-estimate. Dark grey and light grey correspond to $\pm1$ and $\pm2$ standard deviations of the parameter estimates, respectively.}
  \label{INVcomp}
\end{figure}

\begin{figure}[H]
\begin{center}
\includegraphics[width=0.8\textwidth]{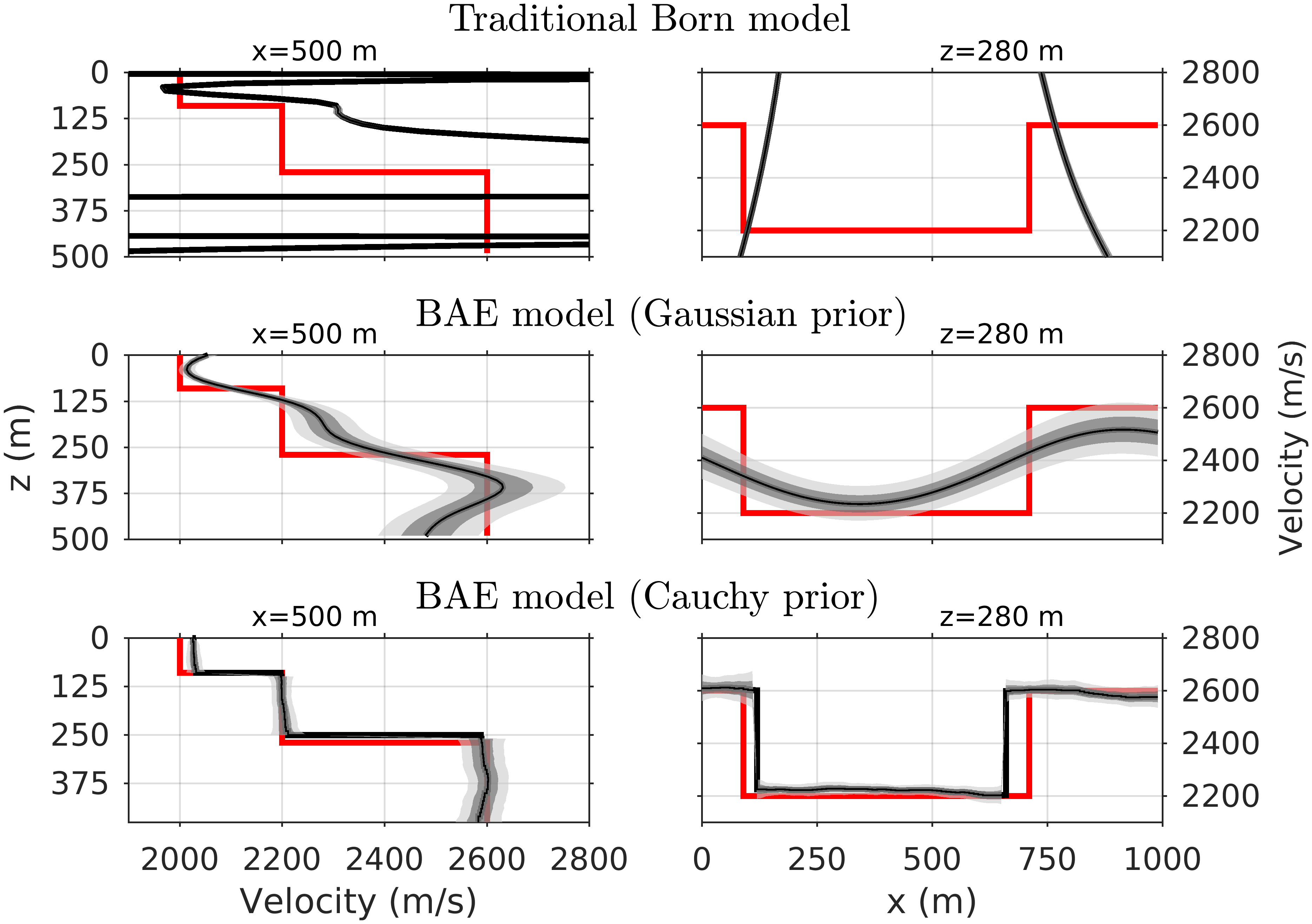}
\end{center}
\caption{Vertical and horizontal profiles of estimated P-wave velocity field for medium B along $x=500$ m (left panel) and $z=280$ m (right panel), respectively: Red lines show the true velocity profile while black lines show the profile corresponding to the CM-estimate. Dark grey and light grey correspond to $\pm1$ and $\pm2$ standard deviations of the parameter estimates, respectively.}
  \label{visco2}
\end{figure}

To quantitatively compare the inversion results in Figures~\ref{INVmodel} and \ref{SEGINVmodel}, we plot the vertical velocity profile on the horizontal line $z=280$ m and the horizontal velocity profile on the vertical line $x=500$ m. The profiles of estimated $V_{p}$ field for medium A and B, are shown in Figures~\ref{INVcomp} and \ref{visco2}, respectively. Also shown along these profiles are the confidence intervals that are based on $\pm1$  and $\pm2$ posterior standard deviations of the parameter estimates---the intervals are constructed for the squared slowness perturbations $\mathbf{m}$ as described in section \ref{sec:interval_estimation} and then converted to velocities using equation (\ref{eqn113}).

Clearly, we obtain useful BAE error estimates since the true profiles are within two posterior standard deviations of the reconstructions. This is not the case for the traditional Born estimate without BAE. While the use of the Bayesian framework allows us to have posterior error estimates, these examples demonstrate the need for modelling all types of errors and uncertainties (BAE approach), else the posterior error is either underestimated or meaningless. This comparison also demonstrates the superior potential of using edge-preserving priors in reconstructing the sharp interfaces of the subsurface velocity field. The Cauchy prior makes the subsurface interfaces and geologic edges more precise and sharper in the inversion estimates and keeps the inversion procedure of noisy seismic data robust.

We now discuss several observations that can be made from Figures \ref{INVmodel} and \ref{SEGINVmodel}. By comparing the inverted and the true velocities, we find that noise could slightly reduce the accuracy of the inverted velocities, but the shape of the true velocity field is recovered when using the BAE approach. Without BAE, the reconstruction produces completely inaccurate, unrealistic estimates, and severe artefacts in the velocity field reconstruction. Thus, without including the approximation errors, acoustic Born waveform inversion of viscoelastic data will most likely lead to meaningless results, as confirmed in the above numerical tests.

\section*{Conclusion} \label{Section6}
In this paper, we considered acoustic Born waveform inversion, which is a linearized version of acoustic full waveform inversion based on Born and acoustic approximations. Thus, the study adheres to the limitations associated with these approximations: small velocity contrasts, low frequencies with respect to the scattering domain, absence of multiple scattering effects, and inaccurate predictions of seismic amplitudes. We have suggested and implemented the BAE approach, a method that takes into account the modeling errors induced by these approximations. Our numerical examples suggests that the BAE approach would allow the use of the acoustic Born waveform inversion of seismic data for viscoelastic media, while neglecting the related modeling errors results in very poor recovery of the subsurface velocity field. Future tests will focus on applying this approach to field data, which is inherently viscoelastic, using non-linear full waveform inversion.  

\segsection*{ACKNOWLEDGMENTS}

This work has been supported by the strategic funding of the University of Eastern Finland and the Academy of Finland (the Finnish Centre of Excellence of Inverse Modelling and Imaging), and Academy of Finland projects 326240, 326341, and 321761. 

\bibliographystyle{seg} 
\bibliography{refs}

\end{document}